\documentclass[aps,prl,twocolumn]{revtex4}
\usepackage{graphicx}
\usepackage{epsfig} 
\newcommand{\bra}[1]{\langle {#1} |}
\newcommand{\ket}[1]{| {#1} \rangle}

\begin{document} 
\bibliographystyle{revtex}
\title{Superconducting and charge-density wave instabilities in
ultrasmall-radius carbon nanotubes}
\author{Ryan Barnett, Eugene Demler, and Efthimios Kaxiras}
\affiliation{Department of Physics, Harvard University, Cambridge MA 02138}
\date{\today}

\begin{abstract}
We perform a detailed analysis of the band structure, phonon dispersion, and
electron-phonon coupling of three types of small-radius carbon nanotubes 
(CNTs): (5,0), (6,0), and (5,5) with diameters 3.9, 4.7, and 
6.8~\AA~respectively. 
The large curvature of the (5,0) CNTs 
makes them metallic with a large density of states at the Fermi energy.  
The density of states is also strongly enhanced for the (6,0) CNTs
compared to the results obtained from the zone-folding method.  For the
(5,5) CNTs the electron-phonon interaction is dominated by the in-plane
optical phonons, while for the ultrasmall (5,0) and (6,0) CNTs the
main coupling is to the out-of-plane optical phonon modes.  We calculate
electron-phonon interaction strengths for all three types of CNTs and
analyze possible instabilities toward superconducting and charge-density wave
phases.  For the smallest (5,0) nanotube, in the mean-field approximation 
and neglecting Coulomb interactions, we find that the charge-density wave 
transition temperature greatly exceeds the superconducting one.  When
we include a realistic model of the Coulomb interaction we find that the 
charge-density wave is suppressed to very low temperatures, making
superconductivity dominant with the mean-field transition temperature 
around one K.  
For the (6,0) nanotube the charge-density wave 
dominates even with the inclusion of Coulomb interactions
and we find the mean-field transition temperature to be around five Kelvin.
We find that the larger radius (5,5) nanotube is stable against
superconducting and charge-density wave orders at all realistic temperatures.
\end{abstract}
\maketitle

%\section{Introduction}

The discovery of carbon nanotubes \cite{Ijima91}
has lead to a renewed interest in the study 
of 1d
electron systems. 
The difference between  semiconducting
and metallic large-radius
nanotubes 
may be typically
understood by 
quantizing the circumferential momentum
of the electronic
states in a single graphene sheet (see, for instance, \cite{Saito98}).
Less conventional properties of nanotubes include Luttinger 
liquid behavior of metallic
nanotubes found in tunneling
experiments (see \cite{Egger00} and references therein),
Coulomb effects \cite{Cobden98}, Kondo physics \cite{Nygard00},
and intrinsic superconductivity observed  in
ropes \cite{Kociak01} and small-radius 
nanotubes in a zeolite matrix \cite{Tang01}.
The main focus has traditionally been on the effects of the Coulomb
interaction between electrons. However, the electron-phonon
interaction has also received considerable attention both
experimentally \cite{Hertel00,Park04} and theoretically 
\cite{Mintmire92, Dubay02, DeMartino03}.
Most theoretical analyses of electron-phonon
interactions in nanotubes assume the phonon frequencies
to be the same as in a graphene sheet and calculate the
electron-phonon coupling strength
from a simplified tight-binding model for the $\pi$ orbitals of the C atoms.
Such an approach, however, may not be suited for ultrasmall
nanotubes (such as the ones in \cite{Tang01}), for which 
the curvature of the nanotube
leads to strong hybridization of the $\sigma$ and $\pi$ orbitals,
which results in a 
qualitatively different band structure \cite{Blase94},
phonon spectrum, and electron-phonon interactions.

In this Letter we present
detailed analysis of three representative small-radius 
nanotubes, the (5,0), (6,0), and (5,5), and discuss possible
CDW and superconducting instabilities of these systems.  
The (5,0) nanotube is the likely candidate
structure for the superconducting behavior seen in \cite{Tang01} with
transition temperature measured around 15 K. The 
radii of the CNTs in this experiment has been determined to 
within 0.2 \AA~by Raman spectroscopy \cite{Li01}.
We demonstrate that even though
standard electronic structure approaches 
for calculating
phonon frequencies,  
such as the frozen-phonon 
approximation (FPA),
run into divergences
intrinsic to mean-field
calculations in 1d,
they can be analyzed from the point
of view of the random-phase approximation (RPA)
for the electron-phonon system and
parameters of the effective Fr\"{o}lich Hamiltonian
can be extracted. The main results that we 
obtain are:
(i)  
The strongest electron-phonon coupling for the (5,0) and (6,0) CNTs 
is to the out-of-plane phonon modes.  This is in contrast to the larger
radius CNTs which have strongest coupling to the in-plane phonon modes
as predicted by the nearest-neighbor tight-binding model~\cite{Saito98}.
(ii)  
Even when the residual Coulomb interaction between electrons is
neglected, the larger radius (5,5) CNT remains stable down to extremely
low temperatures.  For the smaller radii (5,0) and (6,0)
CNTs, when the residual Coulomb interaction is neglected, the CDW 
instability was found to be dominant over superconductivity 
for both types of nanotubes.
(iii)  
We include the residual Coulomb interaction between electrons following
a model developed in Ref.~\cite{Egger98}.  For the (6,0) CNT we 
find that the CDW transition is essentially unaffected by
including the Coulomb interaction and we obtain $T_{\rm CDW}$=5 K.
By contrast, for the (5,0) CNT both CDW and SC are suppressed but now
superconductivity becomes dominant with $T_{\rm SC}$ around 1 Kelvin.

The interaction between conduction electrons and vibrations of
a crystal lattice is commonly described by using the 
Fr\"{o}lich Hamiltonian
\begin{eqnarray}
\label{Eq:frolich}
{\cal H_{\rm e-ph}}&=&\sum_{k\tau \sigma}\varepsilon_{k \tau}
c_{k\tau\sigma}^{\dagger}c_{k\tau\sigma}+
\sum_{q\mu}\Omega^0_{q\mu}(a_{q\mu}^{\dagger}a_{q\mu}+
\frac{1}{2})
\nonumber\\&+&\sum_{k\tau k'\tau' \sigma\mu}g_{k \tau k'\tau' \mu}
c_{k\tau\sigma}^{\dagger}c_{k'\tau'\sigma}(a_{q\mu}+a_{-q\mu}^{\dagger}).
\end{eqnarray}
Here $c^\dagger_{k\tau \sigma}$  creates an electron
with quasimomentum $k$ in band $\tau$ with spin $\sigma$,
$a^\dagger_{q\mu}$ creates a phonon 
with lattice momentum $q$ and polarization $\mu$,
and $q=k-k'$ modulo a reciprocal lattice vector.
The energies of electron quasiparticles and phonons in the absence of
electron-phonon coupling are given by $\varepsilon_{k\tau}$ 
and $\Omega^0_{q\mu}$ respectively
and the electron-phonon vertex is given by $g_{k \tau k' \tau' \sigma \mu}$.

\begin{figure}[htp]
\includegraphics[width=3in]{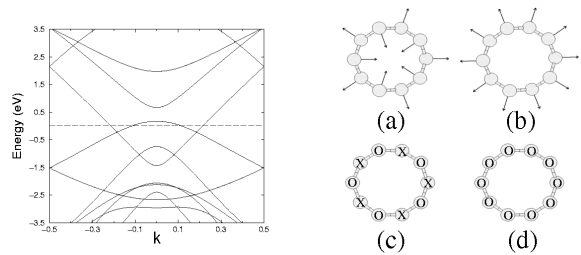}
\caption{Left: The band structure of a (5,0) nanotube where we
set $E_{F}=0$.  Right: The phonon modes that have the
strongest electron-phonon coupling.  Shown is a cross-sectional
slice of the nanotube containing 10 atoms.  Modes (a), (b), (c), and
(d) are out-of-plane optical, out-of-plane breathing, in-plane optical,
and in-plane acoustic modes respectively.  The X's and O's denote vectors
in and out of the page.}
\label{Fig:bands}
\end{figure}

To compute the quasiparticle energies $\varepsilon_{k\tau}$ 
of the representative CNTs, we use
the NRL tight-binding method~\cite{Mehl96} which has been tested 
extensively and
provides accurate results on a variety of materials.  
After fully relaxing the structures with respect to the atomic coordinates,
the band structure is calculated.  We find the band structure predicted
by zone-folding to agree very well with the calculated band structure
of the larger radius (5,5) CNT.  However, for the smaller radius (5,0)
and (6,0) CNTs there was found to be qualitative differences as shown
in Fig.~\ref{Fig:bands}.  While zone-folding
arguments predict the (5,0) nanotube to be insulating, the
band structure clearly exhibits metallic behavior.  The inner band (with
the smaller $k_{F}^{A}$) is doubly degenerate while the 
outer band (with the
larger $k_{F}^{B}$) is nondegenerate where we have
the exact relation $2 k_F^{A} = k_F^{B}$.
The failure of the zone-folding procedure
is due to the strong curvature effects, which
lead to considerable band shifts in small-radius
nanotubes, as discussed originally in ~\cite{Blase94}
for (6,0) nanotubes based on density-functional theory
calculations. As a result of these band shifts, for
the (5,0) nanotubes we have a system close to a Van Hove 
singularity which  has a density of states of 
0.16 states / eV / carbon atom.
For comparison, the density of states for the (5,5) nanotube
is only $0.028$ states / eV / carbon atom, 
so we expect that instabilities of the electron-phonon
systems for the ultrasmall nanotubes are strongly enhanced
compared to larger radius nanotubes.

The electron-phonon vertex given in Eq.~(\ref{Eq:frolich}) can
be expressed as
\begin{equation}
\label{vertex}
g_{k\tau k'\tau'\mu}=\sqrt{\frac{1}{2\Omega^0_{q \mu} M N N_c}}
M_{k\tau k'\tau'\mu}
\end{equation}
where 
$
M_{k\tau k'\tau' \mu}
=\frac{1}{u}\bra{\psi_{k\tau}}(V^{q\mu}-V_{0})\ket{\psi_{k'\tau'}}
$.
Here, $V^{q\mu}$ is the crystal potential under the presence of a 
phonon specified by the ionic displacements
$\delta
{\bf R}_{ni}=u e^{iq R_{n}} \hat{\epsilon}_{q\mu}(i)$
and $V_{0}$ is the crystal potential at equilibrium.
We calculate the magnitude of these matrix elements for 
the coupling between electrons on the Fermi surface to all
phonon modes.  For the (5,0) and (6,0) CNTs we find
that the strongest coupling are to out-of-plane modes.  More specifically,
the strongest overall coupling was found to be to the out-of-plane optical
mode followed by the breathing mode which are shown in Fig.~\ref{Fig:bands}.  
This is in contrast to the larger radius (5,5) CNT 
which has dominant coupling coming from an in-plane optical mode.
We point out that in general, the phonon modes of CNTs cannot
be classified as in-plane or out-of-plane~\cite{Saito98}.  However, the
modes that have the strongest electron-phonon interactions for
the CNTs we study still allow such characterization (see Fig.~\ref{Fig:bands}).
Moreover, CNT dynamical matrix calculations~\cite{Barnett04} show that
the eigenvectors of these modes are essentially the same as in the
graphene sheet.

We point out that, in general, the phonon mode eigenvectors will be
influenced by curvature effects, and will differ from the graphene results.
We checked the relevant modes by using the CNT dynamical matrix, and 
found that they agree well with the zone-folding results~\cite{Barnett04}.

\begin{figure}[htp]
\includegraphics[width=3in]{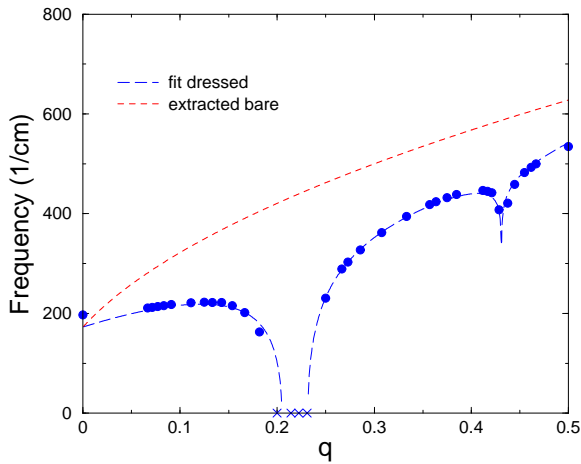}
\caption{The phonon dispersion for the out-of-plane optical
mode in the (5,0) CNT showing 
logarithmic divergences at $2k_{F}$.  We show the fit to the
out-of-plane optical mode for which the
bare frequencies are extracted as discussed in the text.  
The X's denote values of q at which
the calculated FPA frequencies were imaginary.}
\label{Fig:phonons}
\end{figure}

Now we move on to the issue of how to calculate the bare phonon
frequencies $\Omega^0_{q\mu}$ in the Fr\"{o}lich Hamiltonian.  In
the standard FPA \cite{Yin82}, 
the frequencies are given by
\begin{equation}
\Omega_{q\mu}=
\frac{1}{u \sqrt{M N_c}}\sqrt{\Delta E_{\rm cos} + \Delta E_{\rm sin})}
\end{equation}
where $u$ is the amplitude of the displacement, and $\Delta
E_{\rm cos}$ and $\Delta E_{\rm sin}$ are the energy differences per unit cell
between the distorted and  equilibrium lattice structures where the
distortion corresponds
to the real and imaginary parts of $\delta{\bf R}_{ni}=u e^{iq R_{n}}
\hat{\varepsilon}_{q\mu}(i)$ respectively.  The phonon dispersion
curve for the out-of-plane optical mode obtained from the FPA
for the (5,0) CNT is shown in 
Fig.~\ref{Fig:phonons}. Unit cell sizes of up to 400 atoms in
increments of 20 (which is the number of atoms in the smallest
possible unit cell) were used in the FPA, requiring
the phonon wave vectors to be commensurate with the chosen supercell.
This mode shows giant Kohn anomalies 
at $2 k_{F}$ for the inner and outer bands. 

It is important to realize that the divergence
of $\Omega_{q\mu}$  shown in Fig.~\ref{Fig:phonons} does not imply
the divergence of $\Omega^0_{q\mu}$ in the
Fr\"{o}lich Hamiltonian Eq.~(\ref{Eq:frolich}).
In the FPA, 
the phonon frequencies are calculated
{\it after} the electron-phonon
interaction in Eq.~(\ref{Eq:frolich})
have been included, which gives
anomalous softening at $2 k_F$
due to the well-known
Peierls instability of electron-phonon
systems in 1d.
We have developed a technique to extract the bare phonon frequency
$\Omega^0_{q\mu}$ from the numerically computed
$\Omega_{q\mu}$  using a connection
between the frozen-phonon approximation  
and the RPA treatment of the giant Kohn anomaly which is articulated
in Ref.~\onlinecite{Barnett04}.  
Briefly, the dressed phonon frequencies $\Omega_{q\mu}$ 
and the bare phonon frequencies  $\Omega^{0}_{q\mu}$ 
satisfy the equation
\begin{equation}
(\Omega_{q\mu})^2=(\Omega^{0}_{q \mu})^2+2 \Omega^{0}_{q \mu} \Pi_{\mu}(q) 
\end{equation}
where $\Pi_{\mu}(q)$ is the phonon self-energy.
Using the calculated FPA values for  $\Omega_{q\mu}$,
and the calculated electron-phonon coupling values to determine
$\Pi_{\mu}(q)$ (in the random-phase approximation), 
we can extract the bare phonon frequencies 
by assuming that they have the form $\Omega^{0}_{q\mu}=A+Bq+Cq^2$
and performing least-squares fitting where $A,B,$ and $C$ are
adjustable parameters.
Using this method, we have calculated
the bare phonon frequencies of the representative nanotubes, thereby
constructing the effective Fr\"{o}lich Hamiltonians for these systems.

\begin{table}
\begin{tabular}{l|c|c|c|}
&(5,0)&(6,0)&(5,5)\\ \hline
$\lambda_{\rm CDW}$&0.26&0.12&0.024 \\
$\lambda_{\rm SC}$&0.57&0.12&0.031 \\
$\mu_{\rm CDW}$ & 0.24 & 0.0019 & 0.013\\
$\mu^{*}_{\rm SC}$ & 0.19 & 0.16 & 0.093 \\
$T^{0}_{\rm CDW}$ (K)& 160& 5.0& $7\times 10^{-14}$\\
$T^{0}_{\rm SC}$ (K)&64&0.071& $1 \times 10^{-12}$\\
$T_{\rm CDW}$ (K)&$1 \times 10^{-18}$ & 5.0& $2 \times 10^{-43}$\\
$T_{\rm SC}$ (K) & 1.1 & - & - \\
\end{tabular}
\caption{The CDW and SC coupling constants and transition temperatures
for the CNTs studied.   $\lambda_{\rm SC}$ does not include Coulomb 
screening or the
temperature dependent softening of phonons discussed in the text.
$T^{0}_{\rm CDW, SC}$
was computed \emph{without} the residual Coulomb interaction while the residual
Coulomb interaction is included in the calculation of $T_{\rm CDW,SC}$.
$T_{\rm SC}$ includes the temperature dependent renormalization
of $\lambda_{\rm SC}$ (see Eq.~(\ref{lambdaren})).
}
\label{table}
\end{table}

Now we consider possible instabilities of the electron-phonon system.
Using the RPA analysis on the Fr\"{o}lich Hamiltonian, we find 
the CDW transition temperature is given by
\begin{equation}
\label{CDWTC}
T_{\rm CDW} = 4 \varepsilon_{\rm F}
e^{-1/\lambda_{\rm CDW}}.
\end{equation}
Here, $\lambda_{\rm CDW}$ is the dimensionless CDW coupling constant given
by $\lambda_{\rm CDW}=|g_{q \mu}|^2 \nu_{\tau}(0) / \Omega^{0}_{q \mu}$
where $\nu_{\tau}(0)$ is the density of states per spin at the Fermi
energy of the band that is undergoing the transition.  In table \ref{table}
we summarize the results for the CDW coupling constants and transition
temperatures for the nanotubes we study.  The leading CDW instability
for the (5,0) and (6,0) CNTs is from the out-of-plane optical phonon
mode which is shown in Fig.~\ref{Fig:bands} while  the leading
CDW instability for the (5,5) CNT is to an in-plane optical mode.

To study superconductivity, we use the McMillan formula
\begin{equation}
T_{\rm SC}=\frac{\langle \Omega \rangle}{1.20} 
\exp \left[- \frac{ 1.04(1+\lambda_{\rm SC})}{
\lambda_{\rm SC}-\mu^{*}_{\rm SC}(1+0.62 \lambda_{\rm SC})} \right].
\label{TSC}
\end{equation}
where the dimensionless superconducting coupling constant
is given by
\begin{equation}
\lambda=\frac{1}{\nu(0)}\sum_{ k\tau k'\tau' \mu} 
\delta(\varepsilon_{k\tau})
\delta(\varepsilon_{k'\tau'}) |g_{k\tau k' \tau' \mu}|^{2}
\frac{2}{\Omega_{q \mu}}.
\label{BCSLambda}
\end{equation}
In the above equations, $\nu(0)$ is the density of states at the
Fermi energy per spin, $\langle \Omega \rangle=1400$ K \cite{Benedict95}
is the logarithmically averaged phonon frequency, and $\mu^{*}_{\rm SC}$ is 
the Coulomb pseudopotential
which we will set to zero for the time being.
The results for the superconducting coupling constants and transition
temperatures is also summarized in table \ref{table} for the nanotubes
we study.

From this analysis, we find that both the CDW and SC instabilities for
the (5,5) nanotube occur below experimentally realizable temperatures.
This leads one to expect that conventional CNTs of larger
radius also be stable 
down to very low temperatures.  For the
(5,0) and (6,0) CNTs, the CDW instability was found to be dominant
which occurs from coupling to the out-of-plane optical phonon mode.

Now we consider the consequences of introducing the residual Coulomb
interaction. 
We point out that since the charge density is not evaluated self-consistently
in the tight binding method we use, the Hartree term 
which opposes the formation
of the charge-density wave is omitted in our frozen-phonon calculation
of frequencies.  This term
essentially gives the Coulomb energy cost of forming a non-uniform
charge density.  Including the Coulomb interactions properly
should lessen the divergences found at $2k_{F}$ in the phonon spectra.
Introducing the residual Coulomb interaction 
changes our effective Hamiltonian to
\begin{equation}
{\cal H} = {\cal H}_{\rm e-ph}+{\cal H}_{\rm e-e}
\end{equation}
where
\begin{equation}
{\cal H}_{\rm e-e} =\frac{1}{2} \sum_{kk'q\tau \tau' \sigma\sigma'}
V_{q\tau \tau'} c^\dagger_{k+q \tau \sigma} c^\dagger_{k'-q \tau' \sigma'}
c_{k' \tau' \sigma'} c_{k \tau \sigma}
\end{equation}
and ${\cal H}_{\rm e-ph}$ is given by Eq.~(\ref{Eq:frolich}).
For the Coulomb interaction between conduction electrons, we
take the form used by Egger \emph{et al}.~in
Ref.~\onlinecite{Egger98} which, in position space, is given by
\begin{equation}
\label{Eq:EggerCoul}
V({\bf r-r'})=\frac{e^2/\kappa}{\sqrt{(x-x')^{2}+
\left(2R \sin\left(\frac{y-y'}{2R}\right)\right)^{2}+a_{z}^{2}}}.
\end{equation}
Here, the $y$-direction is chosen to be along the perimeter of the
CNT and $x$ measures the distance along the CNT axis.
A measure of the spatial extent of the
$p_{z}$ electrons perpendicular to the CNT is given by
$a_{z} \approx 1.6$ \AA~and $R$ is the CNT radius.  
For the dielectric constant due to the bound electrons, we will take
the value $\kappa \approx 2$ predicted by the model
of Ref.~\cite{Benedict95_2}.

Including the Coulomb interaction in our RPA analysis of the 
CDW instability, we find that the transition temperature is modified
to 
\begin{equation}
T_{\rm CDW} = 4 \varepsilon_{\rm F}
e^{-1/(\lambda_{\rm CDW}-\mu_{\rm CDW})}
\end{equation}
where $\mu_{{\rm CDW}} = \nu_{\tau}(0) V_{q=2k_{F}}$ and $\nu_{\tau}(0)$
is the density of states per spin at the Fermi energy for the band 
that is undergoing the instability.  We thus find that the
effective coupling is directly reduced by including the
Coulomb interaction.  

With our model for the Coulomb interaction,
we find for the (5,0) CNT that $\mu_{\rm CDW}=0.24$ which dramatically
reduces $T_{\rm CDW}$ to around $10^{-18}$ K.  Thus the CDW instability 
for the (5,0) CNT is essentially removed by taking into account the residual 
Coulomb interaction between conduction electrons.  The case is 
somewhat different for the (6,0) CNT, however.  For such metallic 
zig-zag nanotubes, the wave functions at $-k$ and $k$ close to the 
Fermi energy correspond to symmetric and antisymmetric combinations of
atomic orbitals in the graphene 
sheet~\cite{Barnett04}.  Orthogonality of these wave functions
within the unit cell of the CNT leads to the significantly smaller 
$\mu_{\rm CDW}=0.0019$ which essentially does not affect the CDW 
transition temperature.

The residual Coulomb interaction comes up in more subtle ways when
considering the superconducting instability.  By properly dressing the
electron-phonon vertices as well as the phonon propagator in Migdal's 
expression for the electronic
self-energy we find that the renormalized contribution to 
the superconducting coupling constant is
given by
\begin{equation}
\label{lambdaren}
\lambda_{q\mu}=\left(\frac{1}{(1-V_{q}\chi_{0}(q))^{2}}\right)\left(\frac{1}
{1+\frac{2|g_{q\mu}|^{2}}{\Omega_{q\mu}^{0}}\frac{\chi_{0}(q)}{1-V_{q}\chi_{0}(q)}}\right)\lambda_{q\mu}^{0}
\end{equation}
where $\lambda_{q\mu}^{0}$ is the unrenormalized contribution for
a specific process of wave vector $q$ coupling points on the Fermi
surface by phonon mode $\mu$  and 
$\chi_{0}(q)=\sum_{k}(f_{k+q}-f_{k})/(\varepsilon_{k+q}-\varepsilon_{k})$
\cite{Barnett04}.
The first factor describes renormalization of the electron-phonon
vertex by Coulomb interaction and tends to decrease $T_{\rm SC}$
while the second factor corresponds to phonon softening due to the
giant Kohn anomaly in 1d which tends to increase $T_{\rm SC}$.
The temperature dependence described by Eq.~(\ref{lambdaren}) is 
similar to the two parameter RG analysis presented in Ref.~\cite{Grest76}.
In addition to  the renormalization of $\lambda_{\rm SC}$
we also have the direct repulsion between conduction electrons
which is taken into account through the Coulomb pseudopotential
as shown in the McMillan formula \ref{TSC}.
Analysis based on Eq.~(\ref{Eq:EggerCoul}) we find that the inclusion
of $\mu^{*}_{\rm SC}$ eliminates superconductivity in the (5,5) and
(6,0) CNTs \cite{Barnett04}.
For the (5,0) CNT, the main contribution to the renormalization of 
$\lambda_{\rm SC}$ comes the
$2 k_{F}$ coupling to the out-of-plane optical mode discussed earlier.  
Taking into account the temperature dependence in
$\chi_{0}(q)$, using Eqns.~(\ref{TSC}) and (\ref{lambdaren}) we
find a self-consistent solution of $T_{\rm SC}=1.1$ K.  Thus
we see that inclusion of the Coulomb interactions makes superconductivity
dominant over the CDW in ultrasmall (5,0) CNTs.  
We point out that our estimates give a mean-field value of $T_{\rm{SC}}$.
Below this temperature we expect a gradual decrease of resistivity, which
may be described by the Langer-Ambegaokar-McCumber-Halperin formalism
\cite{Langer67, McCumber70}.
Discrepancy between
this value of $T_{\rm SC}$ and the one observed experimentally
of 15 K \cite{Tang01} should
not be a reason for concern because of the exponential dependence
of the superconducting transition temperature on the Coulomb interaction 
strength and the known difficulty in calculating the latter accurately.
For instance, if we replace our estimated value of $\mu^{*}_{\rm SC}= 0.19$
by the commonly used $\mu^{*}_{\rm SC}= 0.10$, we find a 
self-consistent solution for the superconducting transition temperature 
for the (5,0) CNT of $T_{\rm SC}=13$ K.

We thank S. Kivelson, 
I. Mazin, M. Mehl,
M. Tinkham,  and especially B. Halperin 
for very useful discussions.
This work was supported by Harvard NSEC and by
the Sloan foundation.  RB was supported by an NSF
graduate research fellowship.

\end{document}